\documentclass[12pt,a4paper]{article}
\usepackage[utf8]{inputenc}
\usepackage[T1]{fontenc}
\usepackage{amsmath,amsfonts,amssymb}
\usepackage{graphicx}
\usepackage{url}
\usepackage{natbib}
\usepackage{float}
\usepackage{geometry}
\usepackage{tcolorbox}
\usepackage{array}
\usepackage{booktabs} 
\usepackage{multirow}
\usepackage{threeparttable}
\usepackage{setspace}
\usepackage[acronym]{glossaries}

\newacronym{dod}{DoD}{Department of Defense}

\newacronym{esa}{ESA}{European Space Agency}

\newacronym{cnes}{CNES}{Centre National d'Etudes Spatiales}

\newacronym{iss}{ISS}{International Space Station}

\newacronym{eo}{EO}{Earth Observation}

\newacronym{leo}{LEO}{Low-Earth orbit}

\newacronym{nasa}{NASA}{National Aeronautics and Space Administration}

\newacronym{geo}{GEO}{Geostationary orbit}

\newacronym{gnss}{GNSS}{Global Navigation Satellite System}

\newacronym{csa}{CSA}{Canadian Space Agency}

\newacronym{euspa}{EUSPA}{European Union Agency for the Space Programme}

\newacronym{asi}{ASI}{Agenzia Spaziale Italiana}
\newacronym{dlr}{DLR}{Deutsches Zentrum für Luft}
\newacronym{kari}{KARI}{Korea Aerospace Research Institute}
\newacronym{nosa}{NOSA}{Norwegian Space Agency}
\newacronym{nso}{NSO}{Netherlands Space Office}
\newacronym{sso}{SSO}{Swiss Space Office}
\newacronym{uksa}{UKSA}{United Kingdom Space Agency}

\newacronym{espi}{ESPI}{European Space Policy Institute}

\newacronym{bea}{BEA}{Bureau of Economic Analysis}

\newacronym{cba}{CBA}{Cost-benefit analysis}

\newacronym{eumetsat}{EUMETSAT}{European Organisation for the Exploitation of Meteorological Satellites}

\newacronym{insee}{INSEE}{Institut National de la Statistique et des Etudes Economiques}

\newacronym{sirene}{Sirene}{Système Informatique pour le Répertoire des ENtreprises et des Etablissements}

\newacronym{eu}{EU}{European Union}

\newacronym{ape}{APE}{Activité Principale Exercée}

\newacronym{rcs}{RCS}{Registre du Commerce et des Sociétés}

\newacronym{sarl}{SARL}{Société à Responsabilité Limitée}

\newacronym{sa}{SA}{Société Anonyme}

\newacronym{sas}{SAS}{Société par Actions Simplifiée}

\newacronym{naf}{NAF}{Nomenclature des Activités Françaises}

\newacronym{ner}{NER}{Named Entity Recognition}

\newacronym{nlp}{NLP}{Natural Language Processing}

\newacronym{spaas}{SpaaS}{Space-as-a-Service}
\newacronym{sataas}{SataaS}{Satellite-as-a-Service}
\newacronym{ase}{ASE}{Agence Spatiale Européenne}

\usepackage[acronym]{glossaries}
\usepackage{xcolor}
\usepackage{colortbl} 
\usepackage{svg}

\title{A Rule-Based Methodology for Company Identification: Application to the Downstream Space Sector}
\author{Kenza Bousedra\thanks{University of Strasbourg, University of Lorraine, CNRS, BETA. Address: 61 Avenue de la Forêt
Noire, 67000, Strasbourg (FR). Email: k.bousedra@unistra.fr; p.pelletier@unistra.fr} \and Pierre Pelletier\footnotemark[1]}

\date{}

\definecolor{LightGray}{gray}{0.9}
\definecolor{LightSkyBlue}{rgb}{0.53, 0.81, 0.98}
\makeglossaries
\begin{document}
\onehalfspacing
\maketitle

\begin{abstract}
This paper proposes an original methodology based on Named Entity Recognition (NER) to identify companies involved in downstream space activities, i.e., companies that provide services or products exploiting data and technology from space. Using a rule-based approach, the method leverages a corpus of texts from digitized French press articles to extract company names related to the downstream space segment. This approach allowed the detection of 88 new downstream space companies, enriching the existing database of the sector by 33\%. The paper details the identification process and provides guidelines for future replications, applying the method to other geographic areas, or adapting it to other industries where new entrants are challenging to identify using traditional activity classifications.
\end{abstract}

\section{Introduction}
Space activities are undergoing a structural change referred to as 'New Space'. With the digital transformation of the economy, the commercial potential of space activities has increased, offering new market opportunities from space technology exploitation. This implies a new industrial dynamic that affects the entire value chain with new players investing both upstream and downstream and new business models in which space companies' profit comes from the expanded use of space products and services by end users. 

Traditionally, the space economy is divided into two main segments: 
\begin{enumerate}
    \item The \textit{upstream segment} relates to the delivery of space technologies, i.e., satellite, launcher, and ground system design and manufacturing \citep{booz2014, oxfeconomics2009case}. It includes sub-systems, equipment, components, and related software supply \citep{espi2019}. Launch operation services are most often defined as part of the upstream sector \citep{oecd2022}.
    \item The \textit{downstream sector} covers all activities related to the commercial exploitation of space facilities and data to deliver value-added products and services to end-users \citep{morantaifri2022, oecd2022}. Space services are commonly categorized into three application areas: communications (e.g., Direct-to-home broadcasting, satellite radio, broadband Internet), Earth observation or remote sensing (e.g., mapping services, ocean levels monitoring, responses to natural disasters), and navigation (e.g., location-based services) \citep{booz2014, canada2020}.
\end{enumerate}

This work focuses on the downstream segment and the issue of identifying companies actively involved in these activities. As implied by the term 'downstream space segment,' most studies that analyze and evaluate commercial space activities use a value chain representation \citep{canada2020, earsc2021, ukspaceindustry2022, pwccopernicus2019}. This mode of representation requires understanding the various activities and stages that contribute value to a product or service until it reaches the end user. It is the essence of assessment studies that employ the input-output framework to estimate the space sector's contribution to the economy \citep{canada2020, HIGHFILL2022101474}. However, using input-output matrices necessitates selecting sectors that incorporate space activities at the beginning of the analysis. This approach is effective for measuring manufacturing activities but less when applied to value-added services further down the value chain, where activity codes are highly diverse in official industry nomenclatures (e.g., North American Industry Classification in North America, Nomenclature d'Activités Françaises in France). Consequently, studies measuring the size of the space economy in terms of its contribution to industry sectors result in an incomplete downstream segment assessment. 

\bigskip

We propose an alternative approach to identify the players involved in downstream space activities. Understanding the companies participating in the sector would allow us to overcome the problem of explicitly identifying downstream activities within activity codes, a challenge that arises with sector-level analysis. Rather than attempting to estimate the contribution of space within various industry sectors, the idea focuses on detecting the sectors most represented among downstream companies and using this information to identify unknown downstream space businesses. Moreover, we believe knowledge of companies involved in downstream space activities is fundamental in the context of New Space. We define New Space as a paradigm shift in the space sector, marked by the entry of actors intent on developing space applications markets \citep{bousedra2023downstream}. Given this evolving landscape, understanding and monitoring the sector's industrial dynamics is essential when developing an evaluation method for the downstream space segment. It entails detecting the economic actors in the segment and new entrants, analyzing their characteristics, measuring the economic value their activities generate, and following this evolution over time.

\bigskip

This article introduces an identification method based on a text-mining approach. In concrete terms, this method exploits text, a form of unstructured data, to extract new, structured information \citep{tan1999text}, using specific tools that we will detail later. Text mining is increasingly recognized in economic and financial literature as a rich source of information \citep{gentzkow2019text}. It offers the possibility of formulating predictions in various fields, such as stock prices determination \citep{antweiler} or measuring uncertainty in macroeconomic policies \citep{tobback2018belgian}. Recent studies have even exploited the text of corporate websites as an alternative for assessing their economic activity and innovation potential \citep{arora2013entry,gok2015use, libaers2016taxonomy}. Text data mining is not limited to these areas. It can also be used to analyze companies' activities, to propose an alternative industrial classification to the reference classifications \citep{ALHASSAN2013540,hobergphillips, kilephillips}, or even to determine a company's industrial code from its business plan \citep{tongusing}. The textual sources exploited in this approach are varied and encompass scientific articles, speeches, web pages, publications on social networks, or emails \citep{hassani2020text}.

\bigskip

In the context of our research, the information we seek to extract from the text is the names of companies registered in France and involved in downstream space activities. For this purpose, our primary textual source is the French-language newspaper press. There are several reasons for this choice. Firstly, newspapers are a rich textual source in terms of data volume. The digital version of the press offers abundant content, which is particularly useful in text mining and machine learning methods that require large amounts of training data. The second argument is that the press is an effective source of information and economic analysis. For example, it is used to measure the state of an economy \citep{bybee2021business} or to gather information on fraud in finance literature \citep{miller2006press}. Thirdly, the digital format of the press is a rapid information medium, so certain players such as investors consult it in near-real time \citep{hong2002knowledge}. In this way, the use of the press fits in particularly well with our exploratory approach, the aim of which is to discover new companies in an evolving sector that are unknown to industry experts.

\bigskip

We propose to apply a method for identifying named entities to press data called \Gls{ner}. This approach refers to '[the extraction] of specific 'proper nouns' from unstructured texts, such as the names of people, places, and institutions, in addition to dates' \citep{liu2022overview}. In our case, we are dealing with organizations with downstream space activity. Named-entity recognition is a \gls{nlp} task that enables machines to read, understand and interpret human language \citep{gorinski2019named}.  Thus, the \gls{ner} follows a two-step process: it detects the entities cited in a text, then classifies, or \textit{labelise} them according to categories (place, person, organization). \cite{maurya2022building} conducts research very similar to ours, both from a methodological point of view and in terms of the field analyzed. Indeed, the study is based on \gls{ner}'s tool for detecting satellite names, rocket names, and space agencies from Wikipedia and Google news texts. They use a machine learning model, i.e., a system pre-trained to annotate entities in a text correctly.

\bigskip

There are two possibilities for identifying entities in text in \gls{ner}: a statistics-based approach and a rule-based approach. Unlike \cite{maurya2022building}, which employs the statistical method, we use the rule-based approach. In other words, we establish a word dictionary and formulate a series of instructions (or rules) to apply to the press text to match the names of companies involved in downstream activities. Because of their diversity, company names are particularly challenging to recognize compared to personal names. Indeed, company names can be common nouns used as proper nouns, surnames, or first names. It makes it particularly difficult to capture this type of entity using a purely statistical model, and context is essential to limit the over-identification of words like company names in a text. The rules-based approach therefore enables industry experts to formulate customized detection rules.  

\bigskip

In this paper, we provide a detailed description of our \gls{ner} rule-based approach for detecting downstream company names and present the results of the first application of the method to the case of the French downstream sector. The remainder of the paper is organized as follows. In the next section, we provide a general presentation of the method, detailing the main steps and the textual data collected. As our methodological tool is intended to be replicated, we detail in Section 3 the query that enabled us to collect articles on the subject of downstream space activities and formulated rules. In Section 4, we present the main results of the first application of the method and detail the number of downstream space companies it enabled us to discover. Section 5 evaluates specific aspects of this first version of the method. Finally, with the methodological aim in mind, we address the question of the time required to implement the method and its replication. We conclude the paper with a discussion, reviewing the main contributions of our approach and suggesting avenues for improvement.

\section{General overview of the rule-based approach for downstream space company names identification}

\begin{figure}[h!]
\centering
  \includegraphics[width=\textwidth]{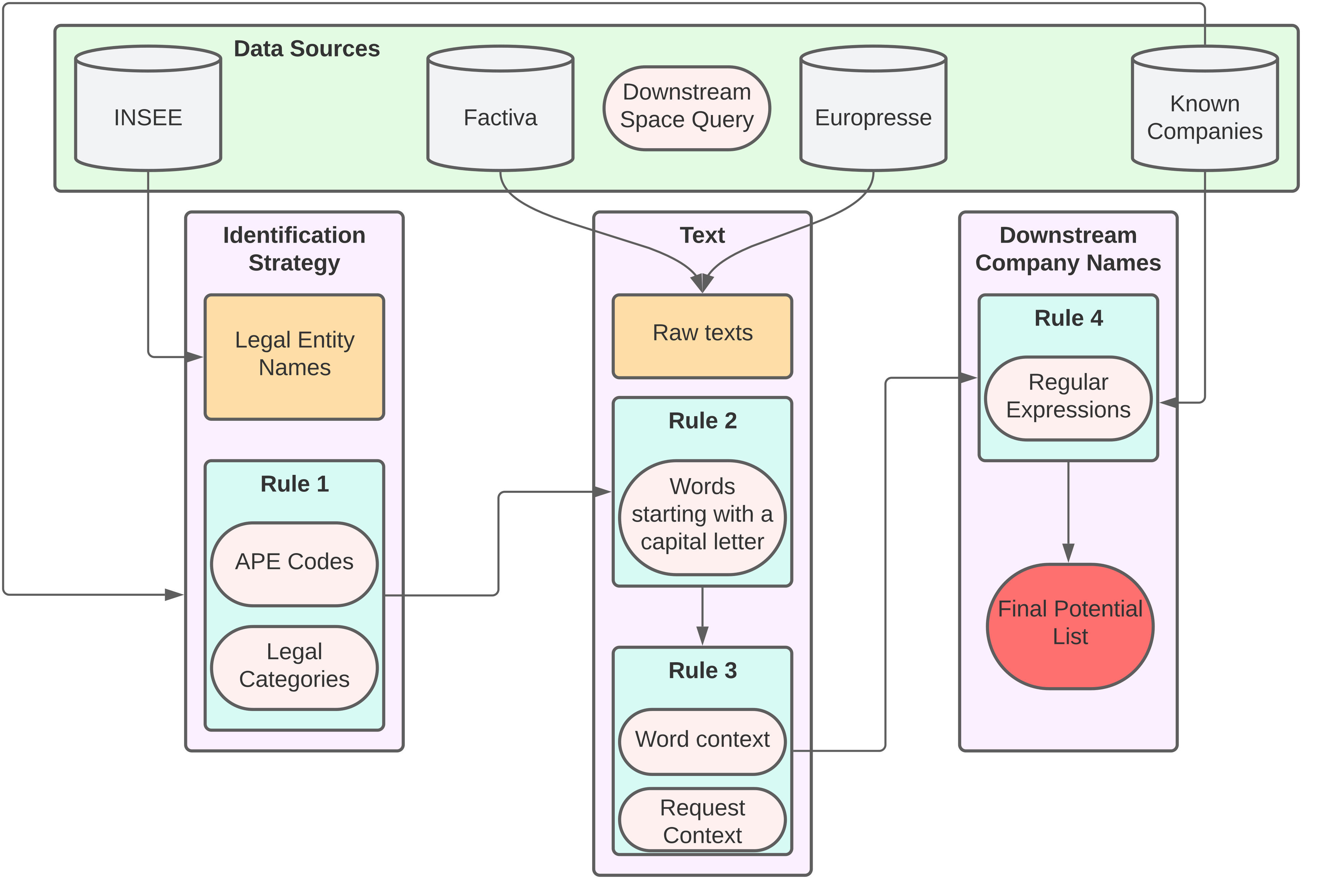}
  \caption{Rule-based strategy}
  \label{figure-method-pres}
\end{figure}

This section comprehensively describes our methodology for identifying French downstream space companies. The proposed approach is designed to supplement internal databases with companies previously unknown to space actors and to overcome the traditional statistical tools' limitations discussed in the introduction. We adopt a \textbf{rule-based method} based on textual information from press databases. The general principle is to extract from news about the French downstream space sector a relatively compact list of companies engaged in downstream space activities. We implement a series of hand-crafted rules to recognize company names associated with the sector within a raw text of information on the downstream space sector. Figure \ref{figure-method-pres} summarizes the critical stages of our approach.

\subsection{Data collection}

The textual data were extracted from four data sources. The first two sources are the electronic press databases Factiva and Europresse, which we used for article collection. Factiva was our primary resource. It is a business-oriented platform offering access to more than 30,000 international sources, including newspapers, journals, business magazines, and newswires. We used Europresse as a supplementary source. The coverage of Europresse is European publications. We used this database to enrich our data with French sources not included in Factiva, such as \textit{Le Monde}, \textit{Libération}, and 89 other sources, including regional publications. The most prominent journals are in Appendix \ref{appendix_info_journaux}, Figure \ref{figure-journals}.

\bigskip

The third database, referred to as the \textit{\acrshort{sirene} database} throughout the paper, encompasses all French-registered companies. This registry, provided as open data by the French National Statistics office (\acrshort{insee}), includes variables like company name, identification code, creation date, and main activity. We selected the legal units file, which refers to the firm as a legal entity, rather than the establishment file, with our primary interest being company names. This legal unit file contains a variable called 'Dénomination usuelle' for names.
The \acrshort{sirene} file is updated monthly, meaning a company registered in a particular month, \textit{m}, will be included in the file for the following month, \textit{m+1}. We used the June 2022 version of the file named 'Fichier StockUniteLegale'\footnote{The file is available at: \url{https://www.data.gouv.fr/fr/datasets/base-sirene-des-entreprises-et-de-leurs-etablissements-siren-siret/}}. This database housed 23,414,852 records, with 13,766,039 corresponding to active entities. Among these active companies, we identified a total of 6,961,344 unique names.

\bigskip

The final database we used, referred to as the 'known companies' database in Figure \ref{figure-method-pres}, contains the names of 220 French-registered downstream space companies. We assembled this database from several different information sources.
Firstly, we leveraged a database on the French downstream ecosystem from a 2016 study conducted by a consulting firm on behalf of \acrshort{cnes}. Secondly, we gathered information on downstream space companies that were recipients of the first space section call of the France Relance program in 2021\footnote{France Relance is the economic recovery plan launched by the French government in the wake of the Covid-19 health crisis. The results of the call for the space sector are available at: \url{https://www.entreprises.gouv.fr/fr/actualites/crise-sanitaire/france-relance/france-relance-premiers-laureats-du-volet-spatial}}.
Thirdly, we sought out information on space companies involved in incubation programs led by \acrshort{cnes}, specifically the \textit{CONNECT by CNES} initiative, which the French space agency launched to support space-related businesses\footnote{The information is available here: \url{https://www.connectbycnes.fr/en/space-for-good}}.
Lastly, we gathered company names from the history of the \textit{ESA BIC} incubation program run by the \acrfull{esa}\footnote{The list of company names is available at: \url{https://commercialisation.esa.int/startups/} Country: France, Space Connection: Downstream}.\\
The known companies database served as a solid foundation for our method because it includes companies definitely involved in downstream space activities. Consequently, we relied on the characteristics of these known companies (names, industry codes, legal status) to construct two rules for our identification strategy.

\subsection{Method and rules description}

\subsubsection{Extraction of newspaper articles dealing with downstream space activities}

The first feature of our methodology involved extracting a corpus of text specifically related to downstream space activities from the press databases. To achieve this, we developed a Boolean query. This query formulated a topic that enabled us to retrieve news articles on France's downstream space sector that could potentially mention companies within this sector. In essence, this first step aimed to narrow down the textual data sphere, firstly to avoid detecting company names unrelated to our area of interest as much as possible, and secondly, to reduce the volume of text to analyze. We present the \textit{downstream space query}, along with details on its construction and additional restrictions applied (geographical scope, publication period, source removal) in section \ref{section_query}.

\bigskip

\begin{figure}[H]
\centering
\includegraphics[scale = 0.5]{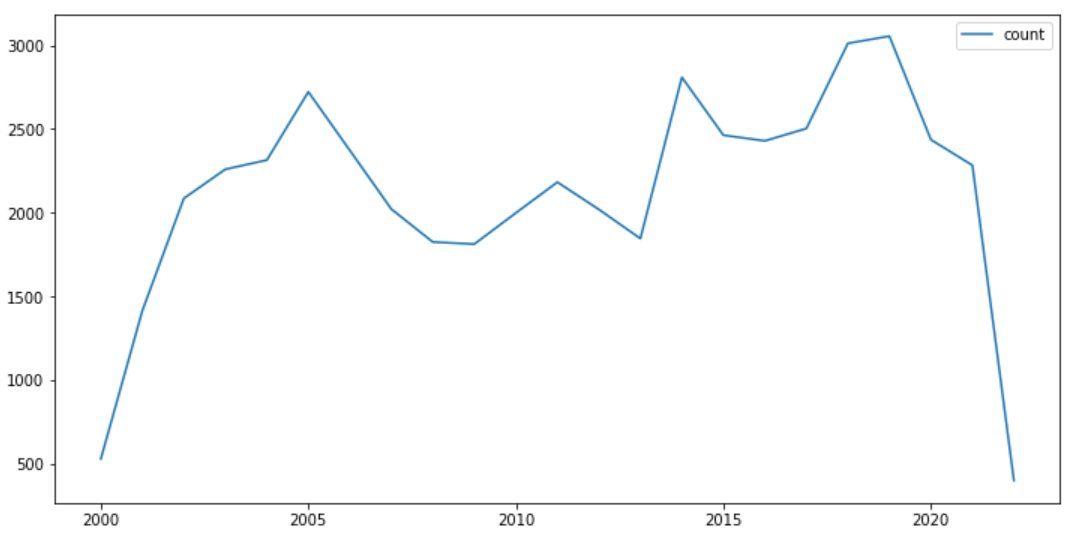}
\caption{Downstream space query results: Number of articles published per year}
\label{figure-article-year}
\end{figure}

The query returned 48,900 articles, 28,400 from Factiva, and 20,500 from Europresse. We manually downloaded the articles in HTML format. We restructured the information they contained (source, author, headline, lead paragraph, body text) into variables using web scraping libraries\footnote{We carried out all the implementation steps with Python software. However, the methodology is replicable with any other programming software.}. The number of articles captured per year ranged between 2000 and 2500 throughout the period (Figure \ref{figure-article-year}).

\bigskip

Once the text was restructured, we eliminated all words that did not start with a capital letter. This critical step forms the \textbf{Rule 2} of the identification method.\footnote{The numbering of the rules does not indicate any temporality in their implementation. Rules 1 and 2 were defined at the same time.} The remaining text contained only punctuation (if a company's name is an acronym or hyphenated), numbers (if a company's name starts with a number), and words starting with a capital letter (e.g., proper nouns and the first word of sentences). From a computational standpoint, this significantly reduced the number of words to compare with the company names from the Sirene database, consequently shortening the processing time. Methodologically, we assumed that since company names are proper nouns, they would always be quoted in the press with an initial capital letter. With this text modification, we minimized the risk of losing company names unless they were incorrectly quoted in the original article.
Furthermore, we observed from the Sirene database that many French company names were very common words. Additionally, news articles often employ non-technical language and simple vocabulary. By eliminating lowercase words from our corpus, we aimed to limit the capture of \textit{false positives}, i.e., a company name that matches a word in the text but does not refer to a downstream company. Finally, we removed all accents to ensure harmonization. 

\subsubsection{Matching the newspaper text with the dictionary of French company names}

The second aspect of the method involved creating a dictionary of company names from the Sirene database. This task required refining the database of French-registered companies to a subset of companies to further limit the number of companies compared with the press text. In doing so, we defined an additional rule (\textbf{'Rule 1'} in Figure \ref{figure-method-pres}): selecting companies with pre-defined APE ('Activité Principale Exercée') codes and legal categories. Limiting the dictionary of company names to business entities belonging to specific sectors of activity allowed us, in the same way as the query with the press articles on downstream activities, to form a list of companies that are part of a field of activity. We selected APE codes and legal categories based on those of downstream space companies in our internal database. We provide the details of the selected codes in section \ref{section_ape}.
It is important to note that a company with a unique national identification number may have multiple names. It is often the case for companies with acronyms or extended full names. The Sirene database provides four name variables ('Dénomination usuelle,' 'Dénomination usuelle 1', 'Dénomination usuelle 2', and 'Dénomination usuelle 3') which we all considered throughout our procedure. Conversely, two active companies with different identification numbers may share the same name. Despite identical names, we maintained two distinct observation lines in our dictionary.

We converted all company names in the dictionary to lowercase and removed accents for consistency. Furthermore, some names in the dictionary included acronyms indicating the company's legal form. We removed these acronyms\footnote{We removed acronyms for Société Anonyme (SA), Société À Responsabilité Limitée (SARL), and Société par Actions Simplifiée (SAS) with and without periods between letters.}, as newspaper articles do not necessarily include the legal form when citing companies.

\bigskip

Next, after removing all lowercase words, we used the Sirene database subset-- our company dictionary-- to detect company names in the articles that came from our downstream space query. We automated this task using a simple algorithm: it compared each company name in the dictionary with each word from the downstream space query text that started with a capital letter. If a word from the article corpus was completely identical to a company name in the Sirene dictionary, it added that company to the list of potential downstream space companies.
Given that this was an exact match operation, both databases compared had to be harmonized (all words in lowercase and without accents). We applied the word comparison to the entire text of each downloaded article (headline, lead paragraph, article body). The result was a list of company names registered in the Sirene database that were cited at least once in one or more articles from our downstream space query corpus.

\bigskip

Though limiting our analysis to capitalized words reduces noise, we still faced instances where we detected words in the Sirene list that corresponded to company names but referred to different entities in the articles. For example, 'Paris' appeared in our list because it was a company registered in the Sirene database and fell within our selected industry codes and legal status. However, many articles mentioned 'Paris' about the city. At this stage in our procedure, we identified Paris as a potential downstream space sector company. We faced this issue with place names, geographical areas, individual names, and acronyms.
Another challenge is that our article corpus did not cover downstream space activities exclusively. An article might mention a downstream space company in one paragraph but discuss topics and companies unrelated to this sector (and the space industry in general) in the rest of the text. Consider an article from a regional publication about a funding plan for local businesses, which includes a company relevant to our research. If we found all mentioned company names in the Sirene dictionary, we would add them to the list of potential downstream space companies. Consequently, using the query as the only filter applied to press databases did not guarantee that we only retrieve text dedicated to downstream space sector information.

\subsubsection{Minimizing False Positives by Considering Citation Context}

To mitigate the incidence of false positives stemming from the above issues, we implemented a third rule (\textbf{'Rule 3'} in Figure \ref{figure-method-pres}) we call the 'Word Context' and 'Request Context' as shown in Figure \ref{figure-method-pres}. This rule relies on two lists of words that establish a context for citing downstream space companies. The first list comprises nouns, verbs, and past participles frequently used in sentences that mention a company name, such as 'start-up,' 'founded,' and 'provide.' By assuming these words are near a cited company in the text, we can lower the likelihood of incorrectly matching names from our Sirene dictionary with unrelated terms in the articles. The second list consists of the main keywords from the downstream space query. We ensure that the matched companies appear within a context that aligns semantically with space activities. This step allowed us to narrow down the company identification to the sections of articles dedicated to downstream space activities. After several trials, we determined a 30-word window for both lists (30 words before and after potential company names). We provide both word context lists in section \ref{section_wordcontext}. For this task, we had to revert to the full text from the query (including lowercase words). The computational time was short since, for each article, we used the list of potential downstream space companies resulting from the initial matching.

\bigskip

By this point in our methodology, we had a list of French companies mentioned in press articles stemming from a query on downstream space activities. We implemented a final, stringent fourth rule directly on this list, called 'Regular expressions' (\textbf{Rule 4}) in Figure \ref{figure-method-pres} and further detailed in section \ref{section_regexpress}. It includes a set of character sequences reflecting as closely as possible the patterns of downstream company names. We chose the patterns that frequently appeared in known companies' database. While this rule may lead to the exclusion of 'true positives' - downstream companies identified by the matching procedure - it aids in eliminating a large number of false positives from the list. We applied the list of regular expressions to the list of potential companies, yielding a final reduced list for manual sorting by an expert. The sorting process involved three steps: First, we verified the company's active status since it might be in the Sirene file but no longer operating. Second, we check the context of its mention in the raw text. This step sometimes enabled us to validate the company name as the article explicitly described the company's downstream activity or discussed an entirely different topic. Otherwise, we checked the company's website to confirm whether it belongs to the downstream space sector based on the company's description.

\bigskip

The subsequent section delves into each facet of the identification methodology we have devised for this research. First, we offer an extensive explanation of the query constructed to compile a corpus of articles discussing downstream space activities. Following that, we outline each of the carefully formulated rules implemented to distill the names of downstream space enterprises from this corpus.

\section{Construction of the query and choice of identification rules}

The development of our method and rules takes into account two primary issues. The first pertains to the method's processing time. Our approach involves an automated part (matching procedures) and a manual part (sorting the final list of potential downstream companies). The objective is to limit the volume of text the computer processes, thus speeding up the procedure and reducing the number of companies to sort. The second aspect relates to the method's performance: the text reduced by the query must be relevant enough to identify as many downstream companies as possible. This optimization of text should confine the \textit{semantic universe} processed to the downstream space sector.

\bigskip

We start by presenting the query and explaining how we developed it. Following that, we describe each rule implemented. Two of these rules apply to the Sirene dictionary (Rule 1 and Rule 4), while the other two apply to press article texts (Rule 2 and Rule 3).

\subsection{\label{section_query}The downstream space query}

The motivation behind building a query is to formulate a search topic within press databases and delimit a perimeter for identifying downstream companies. Its correct formulation is crucial to ensure the performance of the method. Indeed, the matching with the dictionary of company names is performed with the articles from the downstream space query. It must be developed or verified by experts informed on the latest space sector developments, including downstream activities. 

\bigskip

Consequently, we dedicated considerable attention to the formulation and optimization of the query. The more relevant it was, the more likely we were to find companies mentioned in the resulting articles. This step required, on the one hand, the incorporation of words and expressions that covered the activities we were interested in as comprehensively as possible. On the other hand, the query needed to be sufficiently restrictive to exclude articles dealing with activities related to, but not directly part of, the space sector (e.g., non-satellite geographic information) and those dealing with space systems manufacturing. The final query is in Box \ref{fig:box_query}.

\begin{figure}[h!]
\begin{tcolorbox}[colback=blue!5!white,colframe={LightSkyBlue}]
\begin{description}
\item[{\color{cyan} \textbf{(1)}}] 
[satellit* NEAR4 (application* OR service* OR solution* OR operat* OR donnee* OR data OR imag* OR communication* OR telecommunication* OR broadband OR broadcast* OR connectivity OR diffusion OR telediffusion OR cartographi* OR geoinformation OR geo-information OR (information geographique) OR geoloca* OR geoposition* OR geo position* OR position* OR navigation OR surveillance OR monitoring OR tracking)

\bigskip

OR earth observation
OR observation de la terre
OR teledetection
OR remote sensing

\item[{\color{cyan} \textbf{(2)}}] 
OR (downstream space NEAR4 (industr* OR compan* OR provider* OR sector* OR market* OR application* OR service*)) \\
OR (((secteur* OR industrie* OR economie* OR segment* OR ecosysteme*) W/1 spatial*) NEAR30 (aval OR applications))
OR (service* a valeur ajoutee NEAR30 (spatia* OR satellit*))] 
\item[{\color{cyan} \textbf{(3)}}] 
AND (francais* OR french OR france)

\end{description}
\bigskip
\end{tcolorbox}
\caption{\label{fig:box_query}The downstream space query (with Factiva language)}
\end{figure}
 
Our study focuses on companies registered in France. Therefore, we formulate a query with French and English keywords to capture articles from the foreign press in case the method is replicated by including English-language publications dealing with the French downstream market. We developed a single query with words used in English and French, and others specific to each language. However, for this first application of the method, we restrict our search to articles written in French. The addition of English-specific keywords did not change the number of results. 

\bigskip

The query is organized into three distinctive parts. Part (1) in Box \ref{fig:box_query} allows collecting articles referring to activities specific to each application domain (communication, satellite imagery, and navigation). Most of these words taken in isolation do not refer only to space. Therefore, we specified they must not be four words away from \textit{satellit*}. We preferred this term to \textit{space} (or \textit{spatial} in French) to avoid collecting papers on non-satellite geographic data. Two expressions (\textit{remote sensing/teledetection} and \textit{Earth observation/Observation de la Terre}) are used almost exclusively in the space domain and are therefore not subjected to this restriction.  

\bigskip

With part (2) of the query, the goal was to obtain publications that inform about downstream space activities in a more general way. Contrary to the first part, the vocabulary used is different in French and foreign publications. After several attempts and analyses of the database, we confirmed that 'downstream space' is the most widely used expression in articles written in English to qualify activities related to the exploitation of satellite data. In addition, we introduced as a constraint that \textit{downstream space} must not be three words away from \textit{industr* OR compan* OR provider* OR sector* OR  market*  OR  application*} to collect articles tackling the commercial nature of these activities. 

\bigskip

We adopted a similar approach for the French part. The nuance is that the expression \textit{downstream space} (\textit{spatial aval} in French) is not as common as in the English-speaking press. Therefore, we used the same words as in English to obtain articles on commercial space activities (\textit{secteur* OR industrie* OR economie* OR segment* OR ecosysteme*}) followed directly by \textit{spatial*} to specify the domain of activity. Then, we indicated that this expression must not be thirty words away from \textit{aval} or \textit{application*} (i.e., downstream or application) to target the type of activities we were looking for in the space sector. The distance of thirty words ensures that the words are cited in the same paragraph. Finally, we added the expression \textit{service* a valeur ajoutee} (i.e., value-added service) in the same paragraph as \textit{spatia*} (i.e., space) or \textit{satellit*}. It is commonly used in French to describe this type of activity, especially in telecommunications. 

\bigskip

The third part of the query consists in adding as a constraint to the two previous blocks the words \textit{French} (or \textit{Français*}) or \textit{France} in the articles. We tested several options to target articles that cite French companies, such as using the region criterion proposed by the interface. However, introducing the geographic limitation directly into the free text allowed more results. This also left the possibility of using the query on other article databases.  

\bigskip

In addition to the boolean query, we added additional criteria for the Factiva search. We limited the article language to French, thus including foreign press written in French. We removed the sources \textit{EUR-Lex} and \textit{Le Mensuel d'Agefi Luxembourg}. The former is the official journal of the \acrshort{eu}, and the latter is a Luxembourg newspaper dealing with European economic and financial news. Although they are potentially rich sources of information, the articles in these publications were very long. In the first matching tests with the Sirene dictionary, we detected thousands of company names in each paper. In addition to adding much noise, the size of the articles significantly slowed computation time.\\
We started the search with articles published from January 1, 2000, to limit the number of papers to download because this period corresponds to the premises of the New Space. Finally, we removed duplicates, republished news, recurring pricing, market data, obituaries, sports, and calendars. 

\bigskip

We performed the last query in Factiva on February 12, 2022. It resulted in 28,400 articles published between 01/01/2000 and 12/02/2022. We uploaded the HTML pages of the articles.

\bigskip

In addition, we exploited the Europresse platform as a secondary newspaper data source. Our version of Factiva did not give access to the publications of the French daily newspapers \textit{Le Monde} and \textit{Libération}. Moreover, Europresse includes many additional regional sources, particularly those dealing with South-West of France news in which the space sector is established. We used the same query as in Factiva, removing the France restriction (block (3) in Box \ref{fig:box_query}) since we were only looking for articles published in French sources. French articles may deal with another country's downstream space sector news, but we assumed this noise was limited considering the selected sources (mainly regional publications). We also removed the English part in Block (2) and adapted the query formulation to the Europresse language.

\bigskip

As with the Factiva database, we uploaded the HTML pages of the 20,500 articles from the query published between 01/01/2000 and 12/02/2022.

\subsection{\label{section_ape}Rule 1: Industry Codes and Legal Categories} 

The selection of industry codes and legal categories is the \textbf{first rule} of the identification method. We applied it to the Sirene database to create a "Sirene dictionary," which contains only companies with activities and legal categories corresponding to those of known companies. This operation aimed to reduce the number of company names compared with the text from the downstream space query and the computation time.
 
\bigskip
 
Standard industrial classification systems, except for the satellite telecommunications segment, do not allow the identification of firms with a downstream space activity. Most share an activity code with firms unrelated to the sector under consideration. However, we have noticed that specific industry codes were recurrent in our internal database of downstream companies. These are related to information and communication activities, specialized, scientific and technical activities, and business support activities. We therefore applied the list of industry codes from our internal database to the Sirene company database to create a dictionary of companies whose industry codes corresponded to those of known downstream companies. Table \ref{tab:ape} lists the industry codes (\acrfull{ape} codes) selected to build the Sirene dictionary. 

\bigskip

The list includes thirty-three activity codes out of the 732 sub-classes of the French industry classification (\acrlong{naf}). We deliberately filtered the Sirene database with the most detailed code level. Indeed, the \acrshort{naf} is organized into five levels: the section with one letter, the division with two digits, the group with three digits, the class with four digits, and the sub-class with four digits and one letter. Each level provides a more detailed description of the activity of companies in the sector. The more precise the code, the fewer the number of firms that match that code. We selected sub-classes so the rule was restrictive enough, and the dictionary comprised a few companies.

\bigskip

Our code selection covers a wide range of activities. It includes activities related to equipment: the manufacture of navigation and communication equipment (divisions 26 and 30 in Table \ref{tab:ape}) and the sale of equipment (divisions 46 and 47). We naturally kept activity codes related to telecommunications (61 division). Division 52 includes known downstream space companies delivering telecommunication and navigation services in the transportation sector. Division 58 corresponds in our known companies database to downstream companies providing mapping services and software integrating satellite data (e.g., for agriculture). Known companies with an APE code of divisions 62 and 63 provide meteorological services, data processing, and Earth observation services. Divisions 71, 72, and 74 refer to engineering, scientific, and technical activities, but the known downstream space companies in these sectors are very similar to those of divisions 58, 62, and 63. Division 66, referring to insurance activities, stands out from the other selected sectors. One known downstream company with this APE code develops parametric insurance using Earth observation data. 

\begin{table}[h!]
\renewcommand{\arraystretch}{1.1}
\resizebox{14cm}{!}{
\begin{tabular}{ |p{4cm}|p{12cm}| }
\hline
\rowcolor{LightGray}
APE code & Description\\
\hline
26.30Z & Manufacture of communication equipment\\
26.40Z & Manufacture of consumer electronics\\
26.51A; 26.51B & Manufacture of instruments and appliances for measuring, testing and navigation\\
30.30Z & Manufacture of air and spacecraft and related machinery\\
46.51Z & Wholesale of computers, computer peripheral equipment and software\\
46.52Z & Wholesale of electronic and telecommunications equipment and parts \\
46.90Z & Non-specialised wholesale trade \\
47.78C & Other sundry specialized retail sale\\
52.21Z; 52.23Z & Service activities incidental to land transportation; to air transportation\\
58.29B; 58.29C & Development tools and programming languages software publishing; Application software publishing\\
61.10Z; 61.20Z & Wired telecommunications activities; Wireless telecommunications activities\\
61.30Z & Satellite telecommunications activities\\
61.90Z & Other telecommunications activities\\
62.01Z & Computer programming activities\\
62.02A; 62.02B & Computer consultancy\\
62.03Z & Computer facilities management activities\\
63.11Z & Data processing, hosting and related activities\\
63.99Z & Other information service activities\\
64.20Z & Activities of holding companies\\
66.22Z & Activities of insurance agents and brokers\\
70.10Z & Activities of head offices\\
70.22Z & Business and other management consultancy activities\\
71.12A; 71.12B & Engineering activities and related technical consultancy\\
72.19Z & Other research and experimental development on natural sciences and engineering\\
72.3Z  & Data processing \\
74.90B & Sundry professional, scientific and technical activities\\
82.99Z & Other business support service activities\\
\hline
\end{tabular}
}
\caption{\label{tab:ape} Industry codes selected for the Sirene dictionary}
\end{table}

\begin{table}[H]
\renewcommand{\arraystretch}{1.1}
\resizebox{14cm}{!}{
\begin{tabular}{ |p{4cm}|p{12cm}| }
\hline
\rowcolor{LightGray}
Legal Status&Description\\
\hline
3120 & Foreign commercial company registered with the \acrshort{rcs}\\
5499; 5460 & \acrfull{sarl}; Other Cooperative \acrshort{sarl}\\
5599; 5699 & \acrfull{sa} with a board of directors\\
5710 & \acrfull{sas} \\
5800 & European company\\
6599 & Civil company\\
\hline
\end{tabular}
}
\caption{\label{tab:legalstatus}Legal status selected for the Sirene dictionary}
\end{table}

We performed the same operation with legal categories. In other words, we listed the different legal statuses of known downstream companies and filtered the Sirene database by restricting the firms belonging to these legal categories. The list of legal status codes is in Table \ref{tab:legalstatus}. This part of the rule was not very discriminating since most French firms belong to the following three legal categories: \acrlong{sarl}, \acrlong{sa}, and \acrlong{sas}. Nevertheless, it allows for removing all non-profit organizations such as associations, trade unions, and state administrations.

After implementing the rule to the French company database, we obtained a Sirene dictionary that included companies belonging to the selected sectors of activity and whose legal category corresponded to one of the legal codes in Table \ref{tab:legalstatus}. Initially, the Sirene database contained 6.9 million legal units still active. \textbf{The Sirene dictionary after the APE and legal category rule included 650 thousand observations}. The next step of the method was comparing each company name variable of the Sirene dictionary with each capitalized word from the downstream space query. When an exact match occurred, we saved the company name in the potential list of downstream companies. 

\subsection{\label{secton_capitalletter}Rule 2: Words starting with a capital letter}

The third rule also applies to the text resulting from the downstream space query. As described in section 2, it involves keeping only the words from the text that start with a capital letter. We have kept punctuation, symbols, and numbers to avoid losing company names that contain them (for example, acronyms with dots between each letter). This step of the method was also an important source of noise reduction. It prevented a company from the Sirene dictionary with a name also used in everyday language from being matched. For example, the company named 'Sun' could only be matched if the word 'sun' was cited with a capital letter in the text.

\bigskip

Besides, this rule allowed us to reduce the text volume to be compared with each company name from the Sirene dictionary. Therefore, the computation time was lower than if we kept the entire press articles.

\subsection{\label{section_wordcontext}Rule 3: Word Context and Query Context}

The Word Context and Query Context rule \textbf{(Rule 3)} relies on a list of names from business and downstream space activities lexicons. We applied it to the text from the query to reduce the scope of identification of downstream companies to a particular semantic context. During the development phases of the method, we realized that removing only lowercase words in the query text and building up the Sirene dictionary was not efficient enough to substantially reduce the number of false positives. Many companies registered in the Sirene dictionary matched a word in the query text when they were not downstream space companies. We decided to reduce the detection scope of company names in the text. Therefore, we implemented a second matching procedure: We compared the potential list of downstream company names resulting from the first matching with the raw text (with lowercase words). The constraint was the following: names of the potential list must appear in the text within thirty words of at least one of the words of Table \ref{tab:wordcontext} \textbf{or} one of the words of the query (Box \ref{fig:box_query}). The window of thirty was the most efficient in terms of filtering and known companies kept. Generally, a thirty-word window requires that words are in the same paragraph.

\begin{table}[h!]
\renewcommand{\arraystretch}{1.1}
\resizebox{15cm}{!}{
\begin{tabular}{ |p{4cm}|p{12cm}| }
\hline
\rowcolor{LightGray}
Word class & Words\\
\hline
\multirow{8}{*}{\begin{tabular}{l} Nouns and \\ groups of nouns\end{tabular}}
        &   entreprise*, start-up*, startup*, société*, PME, TPE, spin-off*, spinoff*, filiale*, groupe*, pépite*, jeune* pousse*, licorne*, acteur*, spécialiste*, le bureau d’études, opérateur*, fournisseur*, client*, fondateur* de, fondatrice* de, co-fondateur, cofondateur*, cofondatrice* de, co-fondatrice* de,  directeur général de, directrice générale de, dirigeant* de, gérant* de, consortium*, PDG, DG, SA, SAS, SARL, créateur, entrepreneur, incubateur, SATT, pôle de compétitivité, booster, cluster, ecosysteme, business model, modele d’affaires, business plan, capital risque, levée de fonds\\
\hline
\multirow{2}{*}{Verbs}
        &  conçu* par, créé*, fondé*, a développé, développe*, fournit, fournissent, opère*, spécialisé* dans, spécialisé* en\\ 
\hline
\end{tabular}}
\caption{\label{tab:wordcontext}List of context words}
\end{table}

We built the word list (Table \ref{tab:wordcontext}) using a sample of five hundred articles from the downstream space query. In each article citing a downstream company name, we analyzed the context of its citation, i.e., the words preceding and following it in the paragraph. We identified nouns, verbs, and past participles frequently mentioned in proximity to downstream company names. Some nouns refer to the company in general (e.g., \textit{entreprise}, \textit{start-up}, \textit{filiale)} (subsidiary), others refer to company activities (e.g., \textit{opérateur}, \textit{le bureau d'études} (the consulting firm)). The list also contains words that refer to the company management (e.g., \textit{fondateur} (founder), \textit{directeur général} and \textit{PDG} (CEO), \textit{entrepreneur}). In the case where an article cites several downstream companies, we selected words or expressions that refer to a group of companies (e.g., \textit{pôle de compétitivité} (competitive cluster), \textit{ecosysteme}). The verbs often describe the company (\textit{créé} (created), \textit{fournit} (provides, delivers), \textit{spécialisé dans}, \textit{opère}). We included all word forms (masculine/ feminine, singular/plural) in the analysis.

\subsection{\label{section_regexpress}Rule 4: Downstream Regular Expressions}

Once the second word comparison is performed on the text restricted to the semantic context (Word Context rule, section \ref{section_wordcontext}), we obtain a final reduced list of potential downstream company names. However, the list is still too extensive for expert hand-sorting. We introduce a final rule we call Downstream Regular Expressions. It involves selecting only companies detected in the text whose names contain one of the recurring character strings of downstream space company names. This operation is drastic because it considerably reduces the number of companies to sort out by eliminating false positives. However, it also removes true positives, i.e., downstream space companies whose names contain no regular expression. In this respect, we use it with great care: the companies resulting from this rule are only a sample of the total company list where downstream companies are potentially over-represented. It allows us to process a decent number of companies. We therefore do not entirely exclude from the analysis companies that do not belong to this sample.

\begin{table}[h!]
  \centering
\renewcommand{\arraystretch}{1.1}
\begin{tabular}{ |p{8cm}| }
\hline
\rowcolor{LightGray}
Regular expressions\\
\hline
agr, data, e-, farm, geo, ima, lab, map, nav, ocea, sat, sea, service, solution, space, system, tech, tele, terr\\
\hline
\end{tabular}
\caption{\label{tab:regexpress}Downstream regular expressions}
\end{table}

Table \ref{tab:regexpress} lists the regular expressions defined. We exploited the database of known downstream companies to build the selection of character sequences in which we observed recurring identical expressions among company names. Some patterns refer directly to the space sector (\textit{space, sat}), and others give an indication of the application sector of the company (\textit{agr, geo, ima, map, nav, ocea sea, terr}). Finally, some selected expressions are related to company's activities (\textit{data, lab, service, solution, tele}).  

\bigskip

After applying the rules outlined in this section, we obtained a reduced list of \textbf{companies that potentially belong to the downstream space sector}. This list results from the matching between the Sirene company dictionary and press articles on the topic of downstream space. \\
We now move on to the results of the first application of our rule-based identification method.

\section{Results of the first implementation and new downstream space companies identified}

In essence, the first application of our identification method aimed to discover new French downstream space companies, that is, companies that were not known prior to its implementation. In other words, we seek to determine whether the method has successfully enriched the 'known' downstream space companies' database.

\bigskip

Before presenting the final results, let us review the intermediate results, i.e., the number of companies obtained after applying each rule. The application results and the various figures given here are summarized in Figure \ref{evaluation1}.

\bigskip

Initially, we had a file from the Sirene database containing 6.9 million active legal units registered in France. The application of \textbf{Rule 1} (APE Codes and Legal Categories) reduced the number of companies to match with the query text by 90\%, resulting in the \textit{Sirene dictionary} containing 650,000 observations.\\
Next, the list of companies obtained after matching the Sirene dictionary with the query text (only words starting with a capital letter, \textbf{Rule 2}) included 30,084 companies. Thus, this step in the method allowed for a 95\% reduction in the number of companies to sort compared to the Sirene dictionary. \\
To apply the following rule, we remind that we returned to the raw text of the downstream query to determine the citation context of the 30,084 companies obtained in the previous step. By restricting the citation window with the Word Context and Query Context lists to 30 words, we reduced our list of potential downstream space companies to 22,862 observations. \\ Finally, the fourth rule was to keep only company names that contained a 'regular expression.' This rule was drastic, as it allowed us to obtain a final list to sort composed of 1,475 companies.\\
Therefore, our rule-based identification method resulted in a reduced list of companies (from 6.9 million to 1,475)  potentially involved in downstream space activities.

\begin{figure}[h!]
  \centering
  \includegraphics[scale = 0.4]{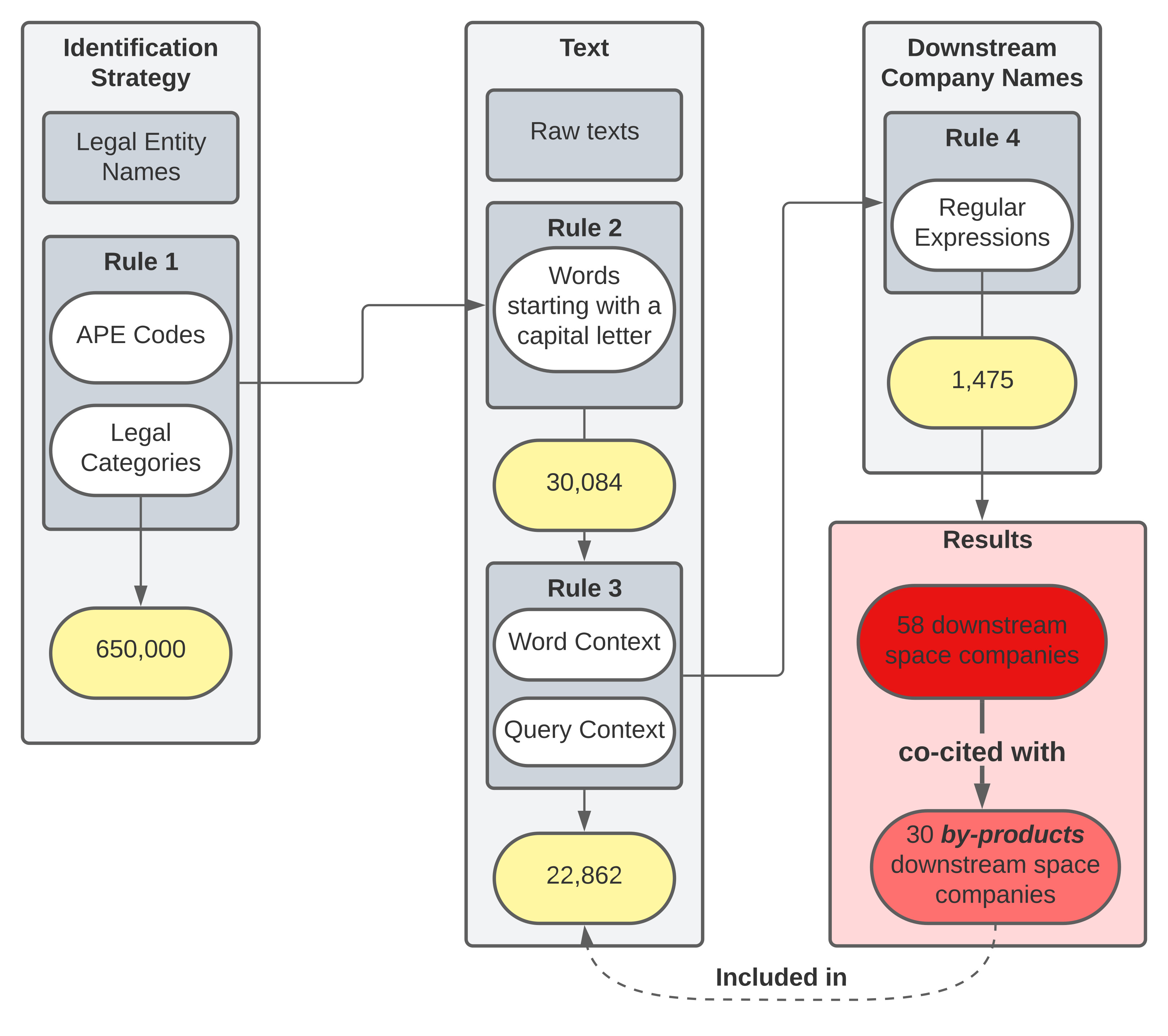}
  \caption{\label{evaluation1}Results of the first implementation}
\end{figure}

The final step in the identification method was to manually sort the 1,475 companies obtained by applying the rules. The procedure for determining whether a company was effectively part of the downstream sector was in two stages: \begin{enumerate} 
\item We returned to the press articles in which these companies were mentioned to see if we could directly classify them as downstream space companies. The sorting procedure stopped here if the article included a description corresponding to a downstream activity.
\item If the citation context in the press was not sufficient, we conducted a search on the company's website. If specific keywords appeared in the description of its offer (e.g., 'satellite,' 'earth observation image,' 'GPS,' 'GNSS'), we classified it as a downstream space company. \\
In some cases, companies were mentioned in application projects or the results of calls for projects led by institutions (CNES, ESA, etc.). This information was also taken into account when sorting the companies.
\end{enumerate}

\textbf{The strict implementation of the method allowed us to identify 58 new downstream space companies.} In other words, 4\% of the companies in the list of company names containing regular expressions were indeed downstream space firms. Given the restrictive rules applied (particularly the last 'regular expression' rule), we find this result satisfactory.

\textbf{In addition, we detected 30 new downstream space companies we call \textit{by-products} of the method.} These companies did not contain a regular expression in their name but appeared in the list of potential companies from the first three rules. During the manual sorting procedure, we noticed papers citing some companies from the list of 1,475 with other companies included in the list of 22,862. Co-citation was of several kinds: companies conducting similar activities or in the same geographical area, participating in the same program or call for projects, or having supplier-client relationships. 

\begin{table}[h!]
\begin{center}
    
\begin{threeparttable}
\begin{tabular}{>{\centering\arraybackslash}m{0.3\textwidth} >{\centering\arraybackslash}m{0.3\textwidth}}
\toprule
\textbf{Source} &\textbf{Number of companies} \\
\midrule
\rule{0pt}{3ex} \raggedright \textbf{Known database} & \\ 
Reference database & 220 \\
Other\tnote{*} & 26 \\
\hline
\rule{0pt}{3ex} \raggedright \textbf{Rule-based method} & \\ 
Method outputs & 58\\
By-products & 30\\
\hline\hline
\rule{0pt}{3ex} \textbf{Total} & 334 \\ 
\bottomrule
\end{tabular}
\begin{tablenotes}
      \item[*] 'Other' refers to companies identified through online research after the method was developed. We will include them in the evaluation stage.
    \end{tablenotes}
\caption{\label{table_summary_downstream_database}Summary of the final downstream space company database by source of identification}
\end{threeparttable}
\end{center}
\end{table}

\textbf{Eventually, we obtained 88 new downstream companies using our rule-based named entity recognition approach.} Table \ref{table_summary_downstream_database} summarizes the final number of downstream space companies that form our database by source of identification. The 'known database' includes the companies we did not identify through our identification method. This database is split into two parts: the 'Reference database,' which served as our basis for formulating the rules, and the 'Other', comprising 26 companies that we discovered via alternate ways (like Internet research and participation in downstream space-focused events) \textbf{post} method application. The 'Rule-based method' column corresponds to the companies that we detected through our approach. Overall, our database of downstream space companies, compiled in 2022, gathers 344 companies, more than 26\% of which were detected using our method.

\section{Calibration and Evaluation of the rule-based identification method}

The identification method presented in this paper results from a series of adjustments and tests to ensure its effectiveness. We started by verifying critical steps of the identification procedure to evaluate their robustness. In the absence of existing studies offering a comparable method for detecting downstream space sector companies, we determined the performance of our process by comparing it to a statistical approach to Named-Entity Recognition.\\

In this section, we first focus on a crucial step of the method, the query formulation. Then, we evaluate the performance of the rules specifically applied to the text and compare it with the statistical approach.

\subsection{Query verification}

To verify the query's ability to obtain news articles that contain downstream space company names, we used our 'Known companies' database. We analyzed how many known companies the query could detect to understand how this step contributes to the overall performance of our approach and where the potential weak points might be for future applications of the method. As a reminder, our method relies not on a machine learning model but on the simple application of rules in newspaper text. Therefore, we have not split our 'known companies' database into a learning and test subset, as our approach has no machine learning process. 

\begin{figure}[h!]
  \centering
  \label{figure-known-comp}
  \includegraphics[width= \textwidth]{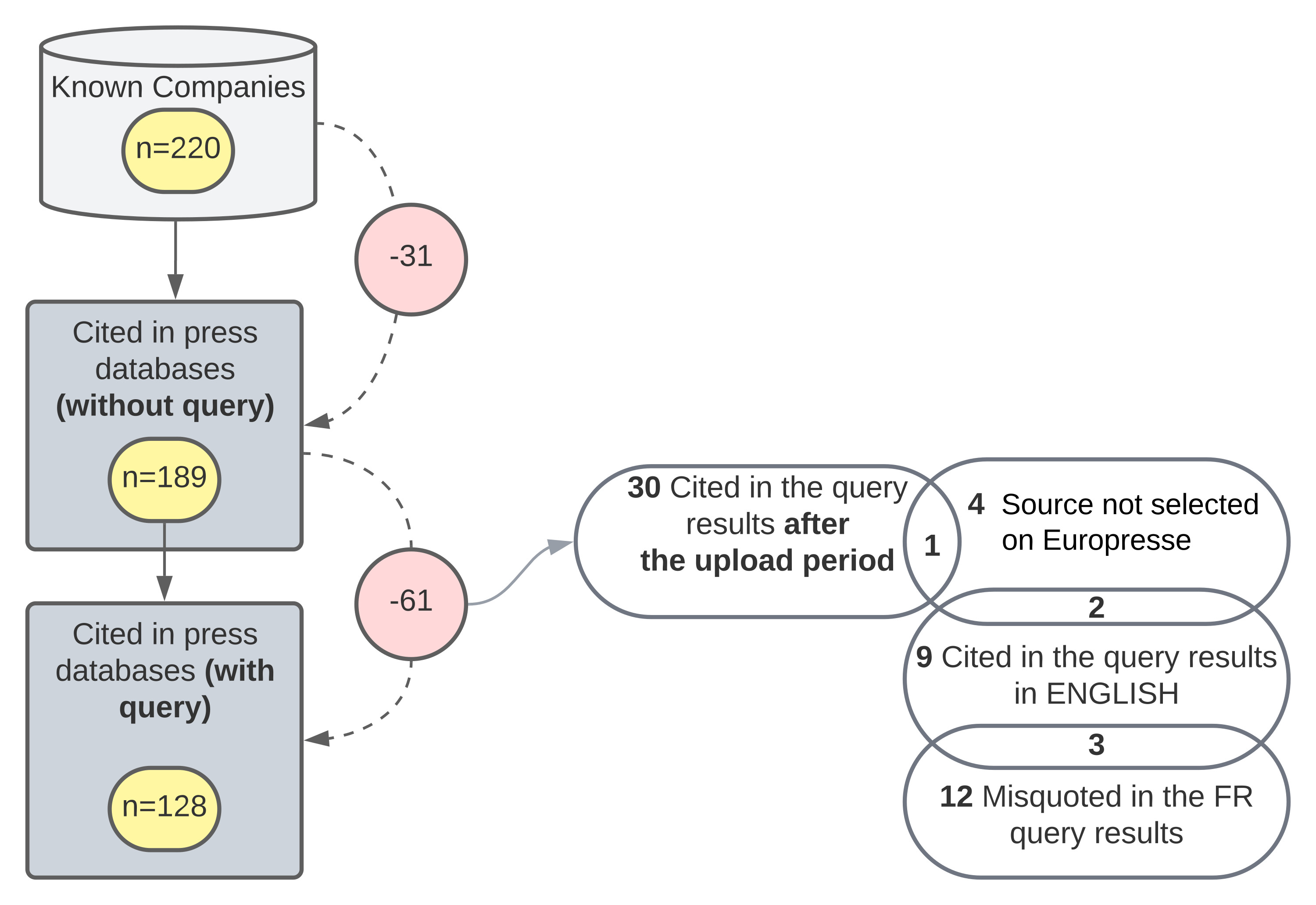}
  \caption{Loss of known companies upon query application}
  
\end{figure}

Figure \ref{figure-known-comp} illustrates the number of known companies before and after the application of the method and indicates the sources of loss after the method's application. The initial known company database included 220 known companies (Figure \ref{figure-known-comp}). The first step was checking whether known companies appeared at least once in the two press databases without applying any restriction. We found that 31 of the 220 companies were neither mentioned in Factiva nor Europresse. In other words, our method could not capture 14\% of known downstream companies since they did not appear in any newspaper text from the databases. Several explanations can be assumed for this. Firstly, these companies were not well-known and did not make the news to appear in the press. Secondly, these companies are still too young and their activity too recent to be the subject of an article.

\bigskip

In the second step, we examined how many known companies among the 189 remaining were cited in the downstream space query results. We expected this part of the method to be an important source of loss since we reduced the text from all available published electronic articles to a corpus restricted by keywords, publication period, and language. Eventually, we found 128 - nearly 70\% - of the remaining known companies mentioned in the papers from the query. Looking at further details on the losses, we observed that half of the companies were not cited in the query keyword context but appeared in the query results \textbf{after} the upload period. It supports our previous hypothesis, namely that there is a delay between the company's creation date and its mention in the press. It is the case for companies that are not part of space programs or not involved in calls for proposals, for example.\\

Misquotes are the second most significant source of losses: 15 known companies were mentioned in the query results but not under the exact name of their legal unit registered in Sirene\footnote{In figure \ref{figure-known-comp}, 12 known companies were misquoted in the French query results. Three were mentioned in the English query results \textbf{and} misquoted in the French query results.}. To illustrate this point, we had the case of a company whose official name, i.e., defined in the Sirene file, was '$X$ France.' This company was mentioned several times in the query but without the word 'France.' Thus, it did not appear in the raw text from the query.
Finally, 14 known companies lost were mentioned in the Factiva query results but only in papers written in English. In addition, we lost 6 known companies by forgetting to select sources on Europresse.

\bigskip

In the following, we detail the part of the evaluation focusing on the formulated rules.

\subsection{Rules performance measurement and comparison with the statistical approach} 

The second part of the assessment relied on the 128 remaining known companies mentioned in the French query results. First, we describe how we assessed the performance of the applied rules in filtering company names relative to how many known firms they identified. Next, we examine the results of applying the spaCy named entity recognition model to our text and compare them with our method. Figure \ref{evaluation1} summarizes the two evaluation procedures.

\subsubsection{Rule performance evaluation}

We applied two rules to the downstream space query text: keep only the words that start with a capital letter (Rule 2) and keep only the words within a maximum distance of 30 words from the Word Context or the Query Context lists (Rule 3). Figure \ref{evaluation1} provides the numbers of companies cited in the text that matched those from the Sirene dictionary, as well as the number of known companies detected after the application of each of the two rules (path 'Rule-based NER,' box 'Evaluation').

\bigskip

We thus measure the performance of each rule by two ratios. The first one evaluates the filtering capacity of the rule. We obtain it by dividing the number of companies after the application of the rule by the number of companies before the application of the rule. The second one evaluates the rule's capacity to detect known companies. We obtain it by dividing the number of known companies detected after applying the rule by the number of known companies before applying it. \textbf{The lower the first ratio (hereafter \textit{Filtering ratio}) and the higher the second ratio (hereafter \textit{Conservation ratio}), the more effective the rule.}

\bigskip

The matching between the Sirene dictionary (Rule 1) and the query text by applying only the rule on capital letters to the raw text (Rule 2) resulted in a list of 30,084 companies potentially involved in the downstream space sector. Initially, the Sirene dictionary contained 650,000 companies. In parallel, applying the capital letter rule and matching the remaining text with the Sirene dictionary almost did not impact the number of known companies identified by the method, with 126 companies detected out of the 128. The two lost companies had names composed of two words, the second written in lowercase. \textbf{The Filtering ratio of Rules 1 and 2 is 0.04, compared to 0.98 for the Conservation ratio.}

\bigskip

Then comes the application of Rule 3, which consists of keeping only the companies mentioned within a distance of less than 30 words from one of the words in the list of Word Context or Query Context (see Section \ref{section_wordcontext}). After applying this rule, we obtained 22,863 potential downstream space companies and detected 120 known downstream space companies. \textbf{Therefore, the Filtering ratio of Rule 3 is 0.76, and the Conservation Ratio is 0.95.} Compared to Rules 1 and 2, the Filtering ratio of Rule 3 is very high, suggesting a low power of filtering. We tried reducing the window size of words ($w=15$ and $w=10$), but we lost too many 'known' companies. It could suggest that the small windows caused us to lose too many companies potentially involved in downstream activities. Moreover, we decided that the company name must be close to one of the words in the Word context list \textbf{OR} Query Context list, which may explain the slight decrease in the number of companies in the potential list. We tested the application of the intersection of the two lists, resulting in a significant loss of known companies.

Overall, Rules 1 and 2 are very effective as they have a significant filtering capacity while preserving the number of known companies. Rule 3, even though its filtering capacity is lower, allowed us to reduce our list of companies to manually sort by nearly 25\% compared to the list obtained after applying Rules 1 and 2.

\begin{figure}[h!]
  \centering
  \includegraphics[scale = 0.4]{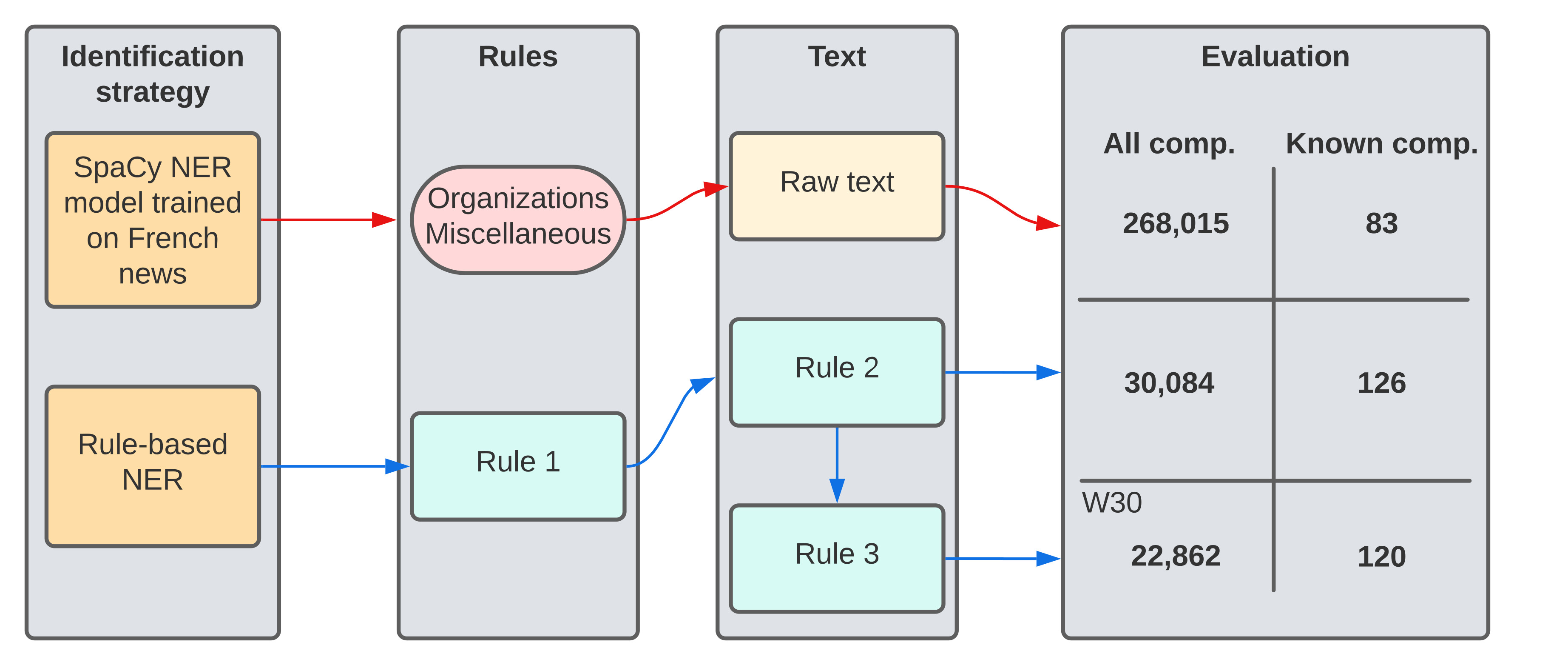}
  \caption{\label{evaluation1}Method evaluation: Rules performance and comparison with the statistical approach}
\end{figure}

\subsubsection{Comparison with the statistical approach}

We compared our results with a statistical approach to have a more general assessment of our rule-based approach. As establishing a set of handcrafted rules to identify downstream space-related companies is time-consuming, we measure the benefits of this method compared to a more straightforward approach that does not rely on explicit rules. The statistical method has the advantage of overcoming spelling inconsistencies as it does not rely on a company names dictionary. Therefore, this approach can recognize company citations in newspaper articles even if their names have been misquoted.

\bigskip

We used \textit{SpaCy}'s pre-trained named-entity recognition model to directly identify companies mentioned in the articles that resulted from the downstream space query (see path 'Spacy NER model trained on French news' in Figure \ref{evaluation1}). \textit{SpaCy} is an open-source Python library designed to simplify complex \Gls{nlp} tasks. In simple terms, \gls{nlp} is an interdisciplinary field between computer science and linguistics, which allows machines to read, understand, and extract meaningful information from human language. \\
SpaCy's functionality is vast. Its applications range from extracting information from text and simplifying text input to more advanced tasks such as interpreting the semantics of a given text. It assists in understanding human language and detecting significant information from raw text. SpaCy provides a highly efficient statistical system for \gls{nlp} in Python, which can assign labels to groups of adjacent words. It provides a default model that can identify a wide range of named entities, including people, organizations, places, and miscellaneous items. In addition to these default entities, SpaCy allows adding arbitrary classes to the model by training it with newly provided examples. 

\bigskip

In our research, we used SpaCy's pre-trained model for the specific task of named entity recognition, with the entity in question being company names. There are many different and efficient \gls{ner} tools \citep{jiang2016evaluating}, but we chose SpaCy as it offers a pre-trained model on French news \citep{jabbari2020french}. Additionally, SpaCy offers a deep learning implementation to obtain dynamic word embeddings, meaning it provides words with a dense vector representation depending on the context\footnote{Documentation is available here: \url{https://spacy.io/universe/project/video-spacys-ner-model}}.

\bigskip

Eventually, in applying SpaCy to our text corpus dealing with downstream activities, \textbf{the \gls{ner} model labeled 268,015 entities as either organizations or 'miscellaneous entities' }(Figure \ref{evaluation1}, box 'Evaluation'). We added the 'miscellaneous' category in addition to organizations to maximize the number of already known companies captured. However, \textbf{only 83 known companies were labeled by the algorithm.}\\
In comparison, our rule-based approach reduced the number of company names to sort to 22,862 and detected 120 known companies out of 128. SpaCy model labeled a significantly higher number of entities but identified fewer known companies than the rule-based approach. This result suggests that our rule-based approach demonstrated a better ability to filter company names and detect known companies. This comparison sheds light on the importance of expert work in constructing an identification method using named-entity recognition tools. In the case of identifying downstream space companies, the rule-based approach and the formulation of customized rules proved more effective than the purely statistical approach.

\section{Time considerations in the application of the method}

One crucial aspect of methodological development that we still need to discuss is the implementation time of the method. As this is a first application, we distinguish here between the time taken to construct the method and the time taken to apply the method. The construction of the method was quite time-consuming as it included the painstaking step of developing customized rules. Before we arrived at the version of the method outlined in this paper, we tested several different rules and went through a series of rule adjustments. The aim was to minimize computation time while maximizing tool performance regarding identification. Thus, the method development phase extended over a year to arrive at the version proposed in this work.

\bigskip

In contrast, the strict application of the method once it was operational was much shorter. The two longest steps were naturally those done manually, i.e., downloading press data at the start of the identification procedure and manually sorting the list of companies resulting from applying the rules at the end of the procedure. For reference, we provide information on the application time per step of the procedure: 

 \begin{enumerate} 
 \item Downloading press articles: 28,400 from Factiva and 20,500 from Europresse.\footnote{We had access to Factiva, which allowed us to download articles in batches of 100 only. On Europresse, we downloaded articles in batches of 1,000.} \\
 \textit{Estimated time: 4 days} 
 \item Matching procedure and rule application (completely automated part of the method).\\ 
 \textit{Computation time: 3 days}
 \item Manual sorting and verification of the 1,475 companies obtained after applying the rules. \\
 \textit{Estimated time: 5 days} 
 \end{enumerate}

Therefore, we estimate \textbf{a total duration of 12 full days for the application of the method}. This duration is variable depending on the context of the application. Suppose it is a first application in another country. In that case, this duration may be longer depending on the number of articles to download and the adaptation of the query if the expert uses other sources for downloading press data. In addition, the list of regular expressions (Rule 4) could change depending on the known database where the method is applied. If it is simply a matter of replicating the method for the French downstream space segment, then this application time could be significantly reduced as the number of articles to download will be fewer (published after 2022 only). If the text to compare with the Sirene dictionary is less voluminous, one can also hope that the list of companies to check at the end of the rule application will be smaller.

\bigskip

We provide recommendations at two levels regarding the frequency and modalities of reproducing the method. Indeed, our methodological developments occur in the particular context of New Space, which we have described as a paradigm shift in the space industry. Thus, we have proposed a tool suited to identifying downstream space companies, given the current specifics of this activity segment. The context is considered in the choice of vocabulary for the downstream space query (also used in Rule 3), in the selection of activity codes for Rule 1 to form the Sirene dictionary, and in the choice of regular expressions with Rule 4. However, new changes may occur in the downstream space sector, with the emergence of new applications in diverse markets and the entry of new-profile companies into the downstream market. In this case, the method will need to adapt to this change.

\bigskip

Hence, the first level of recommendations refers to cases where the expert perceives the downstream space segment as stable compared to the last identification procedure. Without any visible structural change, it involves simply reproducing the identification method to enrich the downstream space database at regular intervals. In this case, \textbf{we recommend reproducing the method annually.} Therefore, the application time is very short since the procedure only includes a year's publications on downstream space activities. In addition, the expert keeps the same formulations for the rules and the query. Therefore, the download, computation, and sorting times are significantly reduced.

\bigskip

However, the expert may notice a critical change in the structure of the downstream segment following a particular shock. For instance, we can imagine the announcement of a large-scale public policy on commercial space applications or a large company investing in developing a new use for space. In this case, some aspects of the method will need reconsideration. The first point is the query formulation, which should be adapted to the new context and the newly emerged downstream activities. Then, the industry codes may need adapting (Rule 1). The expert should analyze the new context and add activity codes to include in the Sirene dictionary. Finally, the rule on regular expressions (Rule 4) might also need reviewing based on the new downstream players. The only stable rule is the one on words starting with a capital letter.

\section{Discussion and areas for improvement}

This paper addressed the issue of identifying companies that are not included in a clearly defined industry within existing activity classifications. Specifically, it focused on identifying new companies involved in downstream space activities. The objective was twofold: to propose a method for detecting companies with downstream space activities and to conduct an initial test of this method. We used Named Entity Recognition tools to extract from a corpus of press text dealing with downstream activities the named entities that we associated, or \textit{labeled}, with the pre-defined category 'downstream space companies'. Our approach was rule-based, wherein we developed a series of customized rules to extract a list of organizations mentioned in the press articles as small as possible and containing the most downstream space companies.

\bigskip

The first rule was to match each word in the text with a Sirene dictionary containing companies with a main activity (industry code) corresponding to that of already known downstream companies. The second rule was to keep only the words starting with a capital letter in the press text, which is generally a characteristic of proper nouns. The third rule involved keeping only the words matched after the previous steps at a maximum distance of 30 words from those appearing in a list of context words or query context. Finally, the last rule aimed to reduce the list of potential downstream space companies by applying a constraint of frequent character sequences, or regular expressions, in downstream company names.

\bigskip

Following the first application of the method, we obtained 88 newly detected downstream space companies. Two-thirds of them were identified through the strict application of the method. The remaining third are 'by-products' of the method and are companies that we identified during the sorting phase of the companies from the list obtained after applying Rule 4. These by-products are co-cited with the companies detected by the method and are in the intermediate list obtained after applying Rule 3. Thus, our database of downstream space companies is composed of 334 companies: 220 companies are known companies used to build the method, 88 companies are companies detected via the rule-based approach, and 26 'other' companies were identified via other sources after applying the method. Eventually, the rule-based approach has allowed us to enrich our database of downstream companies by more than a third.

\bigskip

The proposed approach did not arise \textit{ex nihilo}; we started with a database of known companies and used information about these companies to develop rules and identify previously unknown downstream space companies. Similarly, the method aims to enrich the downstream space database. The tool provides a procedure for regularly updating the downstream space database with companies not yet identified as involved in downstream activities and recently mentioned in the press for this purpose or with newly created, previously unknown companies just mentioned in the press.

\bigskip

A second distinctive aspect of the method is that it heavily relies on expert intervention. This aspect is common when adopting a rule-based approach to NER, as this approach suggests the implementation of directives. The expert is involved in several stages: formulating the downstream space query, elaborating the rules, and sorting the potential downstream space companies' list after matching. Therefore, the method contains both an automated part and a part where the expert's intervention is required. The expert is involved both in applying the method as is and in potential future adjustments we will tackle later.

\bigskip

Lastly, we present a few areas for improvement. One aspect of the method is its heavy reliance on human judgment. Its application is limited to the expert's level of knowledge about the downstream space sector. Additionally, the formulated rules might seem strict and restrictive. The method's ability to capture the downstream space segment's evolution is questioned in this context. The method, as outlined in this paper, could be a starting point toward a completely automated methodology for labeling downstream space companies. Indeed, machine learning algorithms could accelerate the process and regularly propose a list of potential space-related companies to be verified. This way, we could avoid implementing strict rules that are both restrictive and lengthy to establish.

\bigskip

However, as we have seen when comparing the rule-based method to the statistical approach, a method relying solely on the latter would not guarantee results. A potential extension to our work is to merge the two approaches, using first the rule-based approach to build a dictionary of space-specific company names to train a statistical model. For instance, \citep{jafari2020satellitener} trained a model themselves and improved the statistical approach's ability to detect entities related to the satellite domain. They used the spaCy model and predefined three categories related to the satellite domain: organizations, rockets, and satellites. They outperformed state-of-the-art NER techniques to capture items in the Satellite domain.

\bigskip

One aspect to consider is that the diversity of companies related to downstream space activities makes it less direct to construct a dictionary to label a corpus and train a model to capture companies specifically related to downstream space. With only 128 companies detected in our data, it is necessary to feed this list first to increase the number of potential examples of companies related to space. Building another dataset that covers all the texts mentioning these companies is also crucial. Using a reasonably long list of space-related company names and the press articles in which they appear, it is possible to train a model that can recognize space-related organizations and dispense with the rule-based approach in the future.

\bigskip

A last point in our results caught our attention and could be considered for future improvements. We identified a portion of downstream space companies through co-citation. We have referred to these companies as 'by-products' of the method. It would be interesting to delve deeper into this issue of co-citation to establish links between the identified downstream space companies. We only have a raw list of companies for now, but we lack information on potential supplier-customer or competitor links between the companies. Analyzing the co-citation between the known downstream companies and those identified by the method, or the co-citations between the companies identified only, is a potential path to consider in future methodological developments.

\newpage

\printglossary[type=\acronymtype]
\newpage
\bibliographystyle{abbrvnat}
\bibliography{refs}
\section{Appendices}
\subsection{\label{appendix_info_journaux}Information on newspaper texts collected}

Two sources are over-represented in our text database: 'AFP Infos Economiques' with 5,500 articles and 'AFP Infos Françaises' with around 5,000 articles. These two sources are from Agence France-Presse, a national press agency producing dispatches that are then used by other media to relay the information. 

\begin{figure}[h!]
\centering
\includegraphics[width= \textwidth]{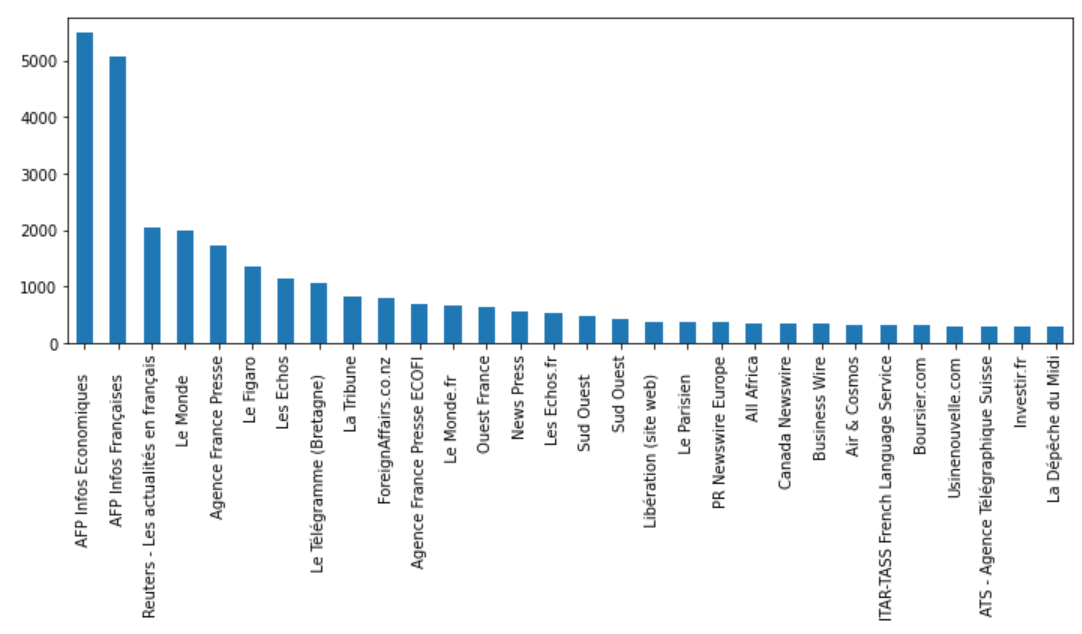}
\caption{Most visible journals, number of articles per source}
\label{figure-journals}
\end{figure}

\newpage

\subsection{Descriptive statistics on the number of companies captured by article}

The density curve below shows a peak at $n=10$, corresponding to the number of companies most frequently detected by our method. 

\begin{figure}[h!]
\centering
\includegraphics[width= \textwidth]{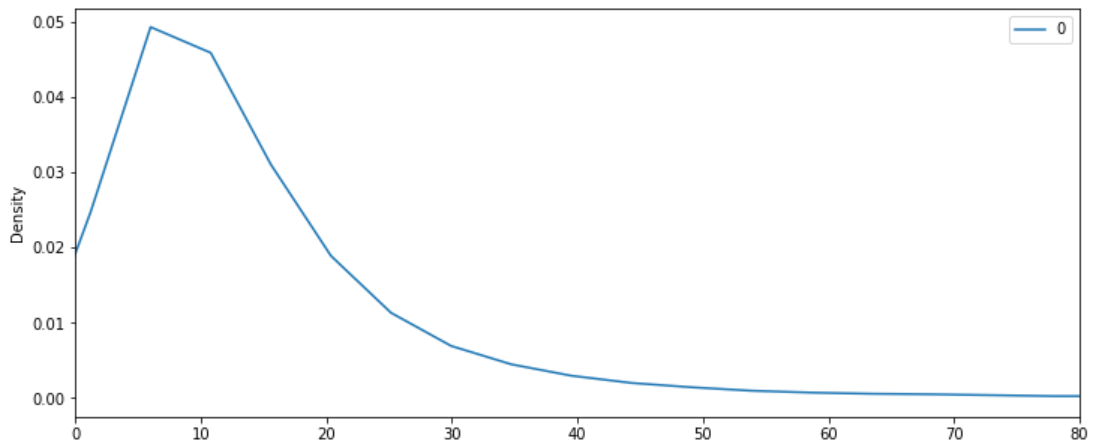}
\caption{Number of companies captured per article}
\label{figure-comp-article}
\end{figure}

\end{document}